\documentclass[mypaper,8pt,twoside]{CoAst}
\usepackage{epsf,graphicx,fancyhdr}
\renewcommand{\chaptermark}[1]{\markboth{#1}{}}

\newcommand{\kopf}{\small\itshape Comm. in Asteroseismology \\ Contribution to the Proceedings of the 38$^{th}$\,LIAC\,/\,HELAS-ESTA\,/\,BAG, 2008
}

\newcommand{\Authors}[1]{\begin{center}\normalsize\bf\sf #1 \end{center}}

\renewcommand{\author}[1]{\begin{center}\normalsize\bf\sf #1 \end{center}}
\newcommand{\Address}[1]{\begin{center}\small\sf #1 \end{center}}

\DeclareGraphicsExtensions{.eps,.jpg}

\newcommand{\Session}[1]{{\vspace{3mm}\small \noindent  \hspace*{3mm} Session: } #1 \normalsize}

	\newcommand{\poster}{\small Poster}

\renewenvironment{abstract}{\section*{Abstract}\normalsize\sf}{}
\newcommand{\References}[1]{\begin{flushleft}{\large References\\}\vspace*{2mm}\small #1 \end{flushleft}}

\newcommand{\chapterCoAst}[2]{\chapter[\sf\normalsize #1\\ \footnotesize \hspace*{5mm}by #2 \sf\normalsize][]{#1\\}\rhead[\fancyplain{}{\sf\footnotesize \center{#1}}]{\fancyplain{}{\sffamily\thepage}}\lhead[\fancyplain{\kopf}{\sffamily\thepage}]{\fancyplain{\kopf}{\sf\footnotesize \center{#2}}}}

\renewcommand{\chaptername}{}

\newcommand{\acknowledgments}[1]{\vspace*{5mm}\noindent  \textbf{Acknowledgments.} #1}

\def\rfr{\smallskip\par\noindent
        \hangindent=7truemm
        \hangafter=1}

\begin{document}
\sf

\chapterCoAst{Testing the forward approach in modelling $\beta$ Cephei pulsators: \\setting the stage}
{A.\,Miglio, J.\,Montalb\'an, and A.\,Thoul} 
\Authors{A.\,Miglio, J.\,Montalb\'an, and A.\,Thoul}
\Address{Institut d'Astrophysique et de G\'eophysique \\
Universit\'e de Li\`{e}ge, All\'ee du 6 Ao\^{u}t 17 - B 4000 Li\`{e}ge - Belgique
}
\noindent
\begin{abstract}
The information on stellar parameters and on the stellar interior we can get by studying pulsating stars depends crucially on the available observational 	constraints: both seismic constraints (precision and number of detected modes, identification, nature of the modes) and ``classical'' observations 	(photospheric abundances, effective temperature, luminosity, surface gravity).
	We consider the case of $\beta$ Cephei pulsators and, with the aim of estimating quantitatively how the available observational constraints determine the type and precision of our inferences, we set the stage for Hare\&Hound exercises.
	In this contribution we present preliminary results for one simple case, where we assume as ``observed'' frequencies a subset of frequencies of a model and then evaluate a seismic merit function on a dense and extensive grid of models of B-type stars. We also compare the behaviour of $\chi^2$ surfaces obtained with and without mode identification. \end{abstract}

\Session{ \poster } \\ 

\section*{Tools}
In order to set the stage for Hare\&Hound exercises, the following three main components need to be defined:
\begin{itemize}
\item {\bf Theoretical predictions:}
The grid of  models we use is  {\sc betadat} (\textit{Thirion~\& Thoul 2006}). Stellar models and adiabatic frequencies are computed, respectively, with {\sc cles} (\textit{Scuflaire et al.~2008a})
 and {\sc losc} (\textit{Scuflaire et al.~2008b}).
The masses considered in the grid span the domain between 7.8 and 18.5 $M_\odot$, a metal mass fraction $Z$ between 0.01 and 0.025, an initial hydrogen mass fraction $X= 0.70$, and four values of the overshooting parameter ($\alpha_{\rm ov}$) in the range 0-0.25. Frequencies of low-order oscillation modes of degree up to $\ell=2$ are computed for main-sequence models.

\item {\bf Observational constraints:} we consider only seismic constraints, i.e. a subset of theoretical eigenfrequencies of a model M0 in the grid. The effects of rotation on the oscillation modes are not considered in this first step, thus all modes are assumed to be axisymmetric ($m=0$). Concerning the degree of the observed modes, we consider the case where $\ell$ is unknown as well the one where $\ell$ is available as a constraint.

\item {\bf Merit function:} In order to compute a merit function at each point of the grid, we use a double optimisation procedure similar to the one extensively adopted in sdB asteroseismology (see e.g. \textit{Brassard et\,al.~2001} and \textit{Charpinet et al.~2005}). For each model in the grid we find the best global match between ``observed'' and theoretical frequencies by using a standard $\chi^2$ merit function. Then we study the properties of good-fit models looking at minima in the $\chi^2$ as a function of the stellar parameters/properties (e.g. location in an HR diagram, mass, central hydrogen mass fraction ($X_{\rm c}$), mean density, \ldots).
\end{itemize}

\section*{First test}
\begin{table}
 \parbox{0.4\textwidth}{
\caption{Theoretical oscillation frequencies of model M0 considered as observational constraints, the uncertainty on the frequencies is assumed to be 0.1 $\mu$Hz.}
}
\parbox{0.4\textwidth}{
\centering
\label{tab:results}
\begin{tabular}{cc}
  $\ell$ & $\nu$ ($\mu$Hz)\\
  \hline
0 & 57.78\\
1 & 58.93\\
1 & 80.29\\
2 & 39.46\\
\hline
 \end{tabular}
 }

\end{table}
We consider as seismic constraints 4 oscillation modes of the model M0 (see Table 1), with frequencies in the typical domain of the pulsation modes observed in $\beta$ Cephei stars. The main parameters and properties defining M0 are: $M=10\,M_\odot$, $X_{\rm c}=0.2$, $\alpha_{\rm OV}=0$, $Z=0.02$, $\log{T_{\rm eff}}=4.34$ and $\log{L/L_\odot}=4.02$.

\begin{figure}
\begin{center}
\parbox{0.5\textwidth}{\centering \Large $\ell$ not known}\parbox{0.5\textwidth}{\centering \Large $\ell$ known}\vspace{0.1cm}
{\includegraphics[width=0.5\textwidth]{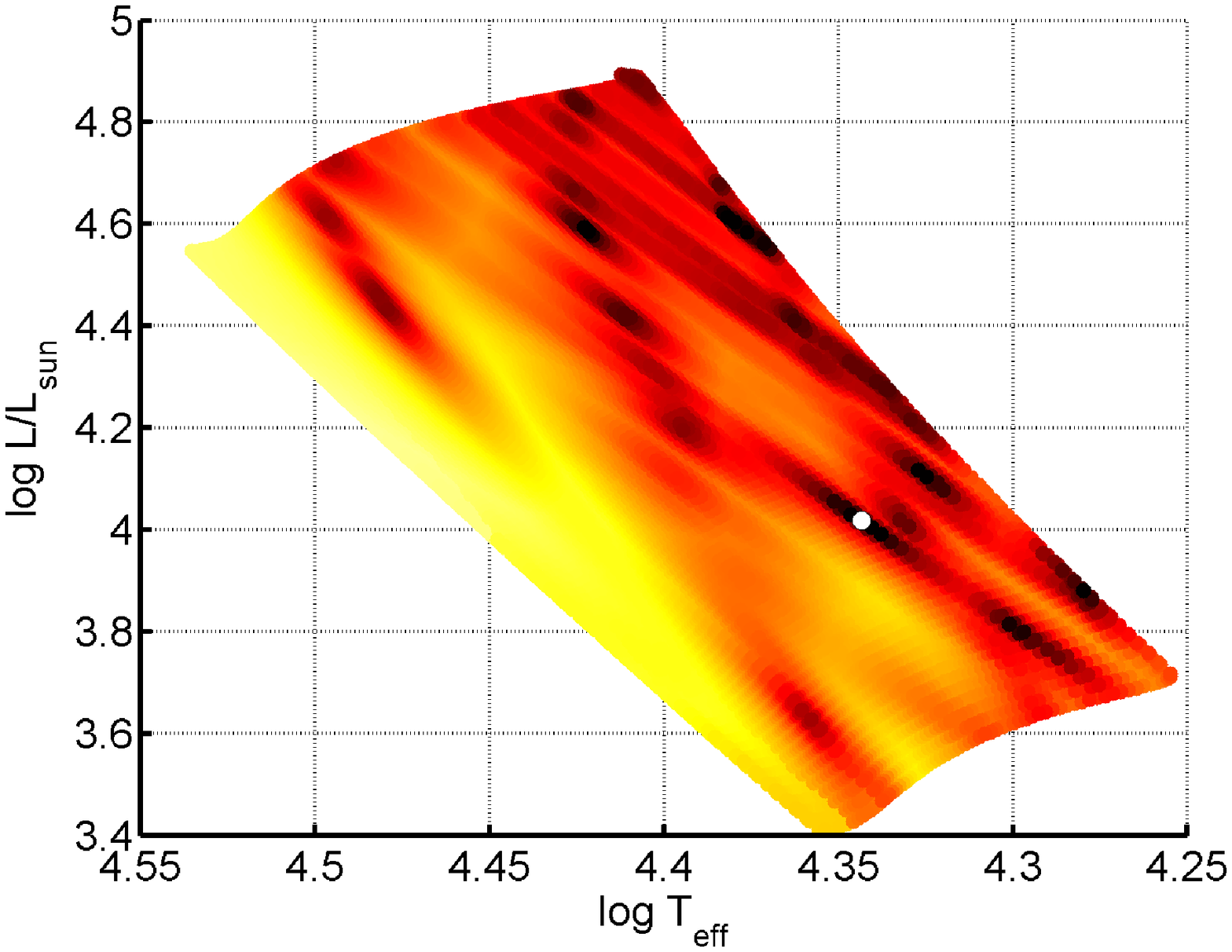}\includegraphics[width=0.5\textwidth]{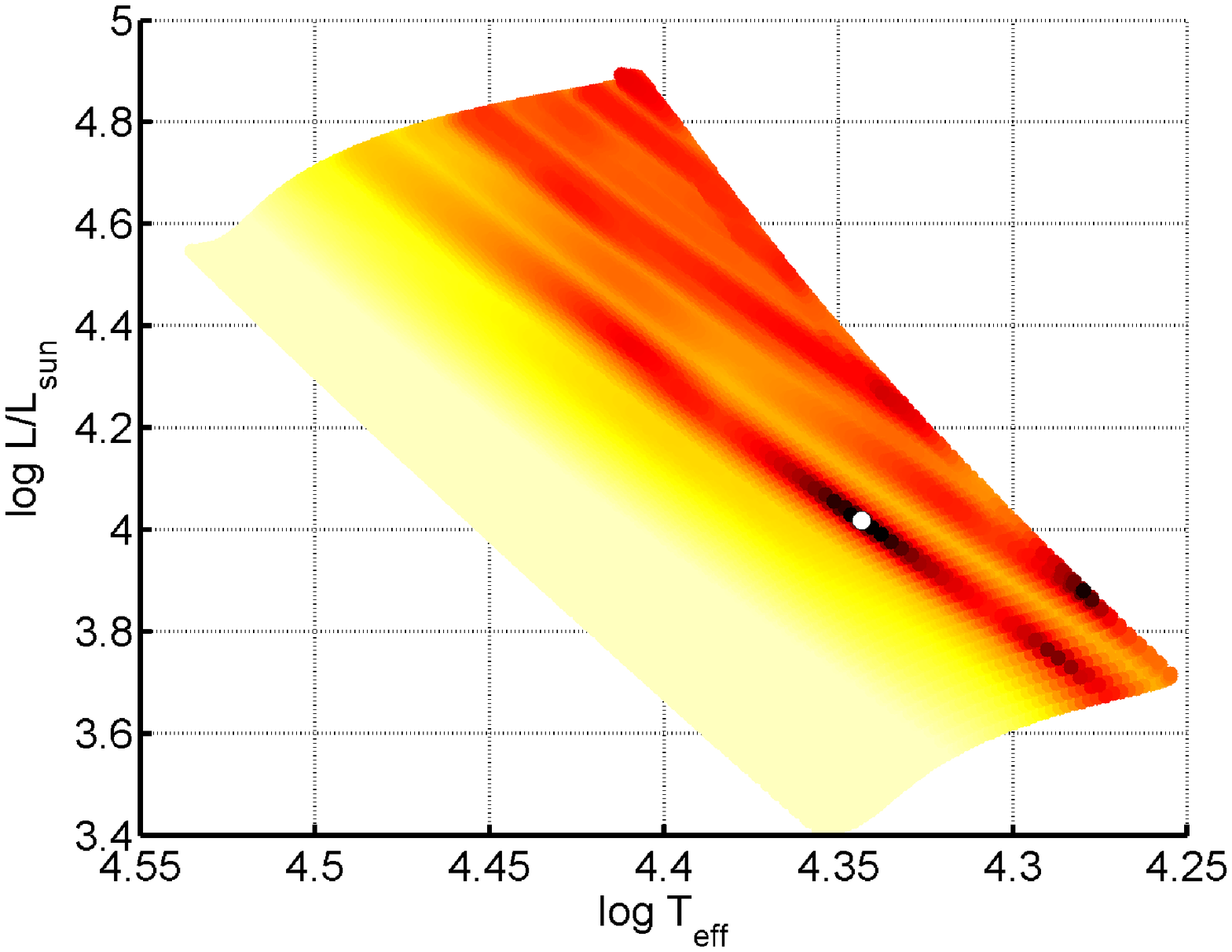}\\
\includegraphics[width=0.5\textwidth]{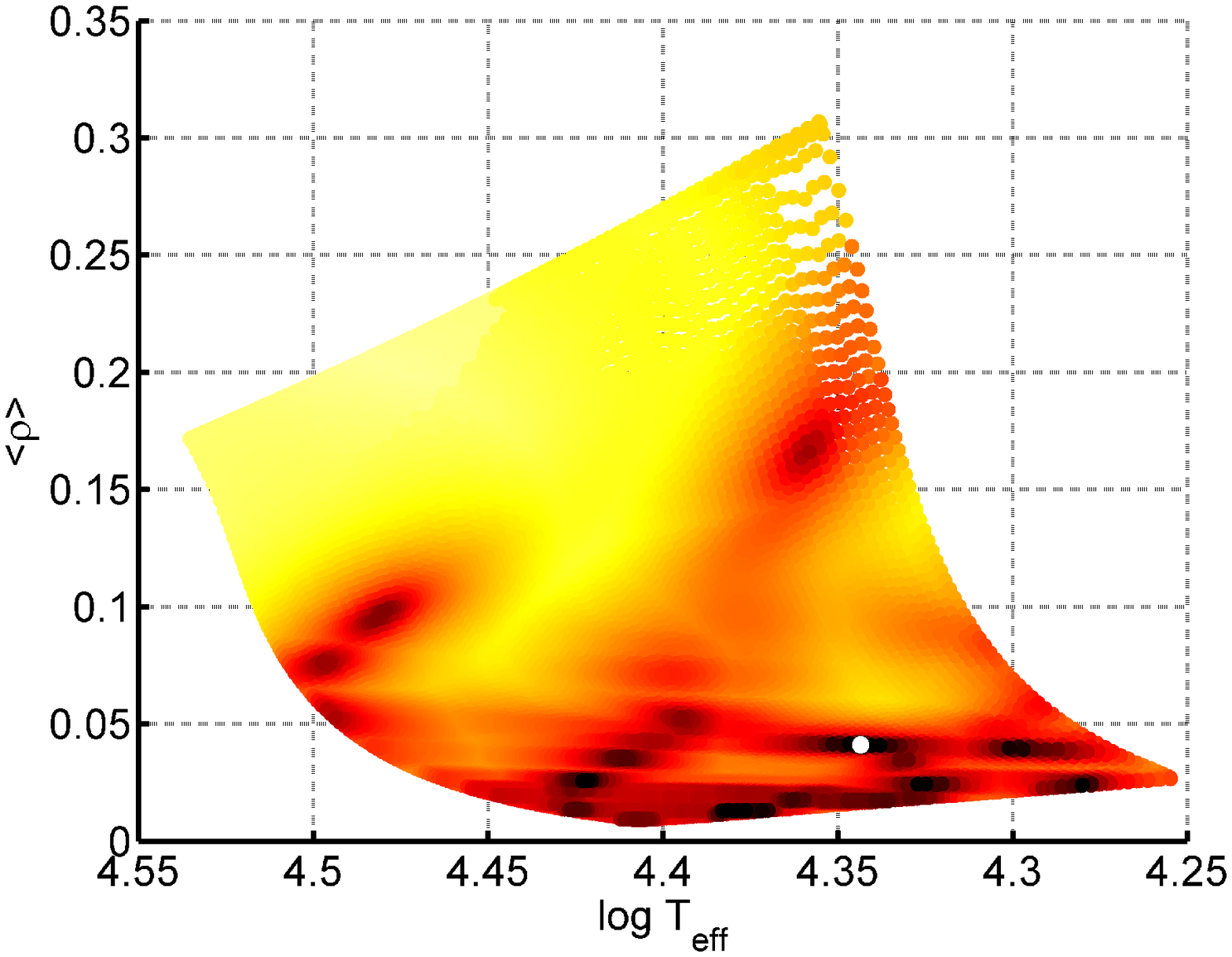}\includegraphics[width=0.5\textwidth]{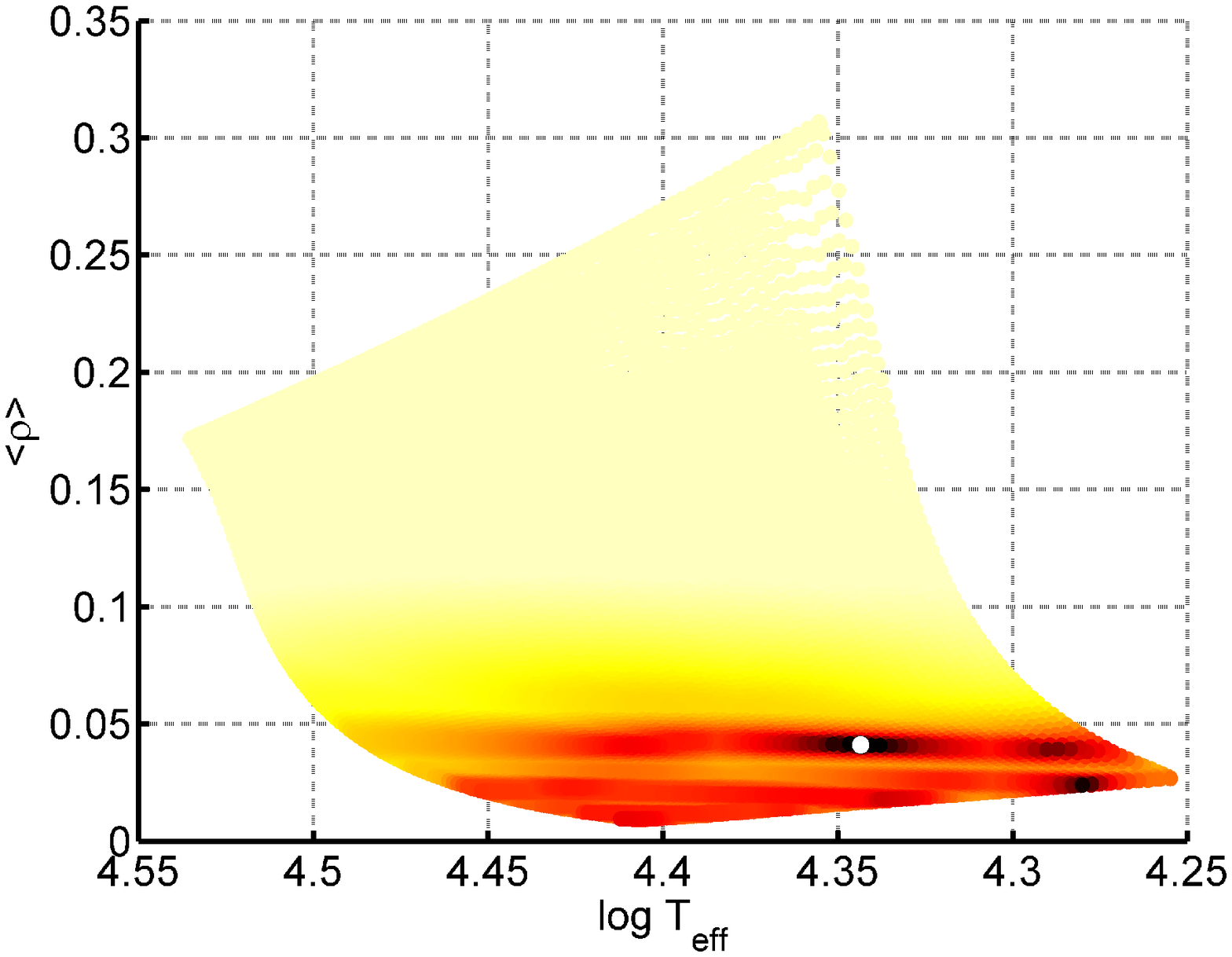}
\includegraphics[width=0.5\textwidth]{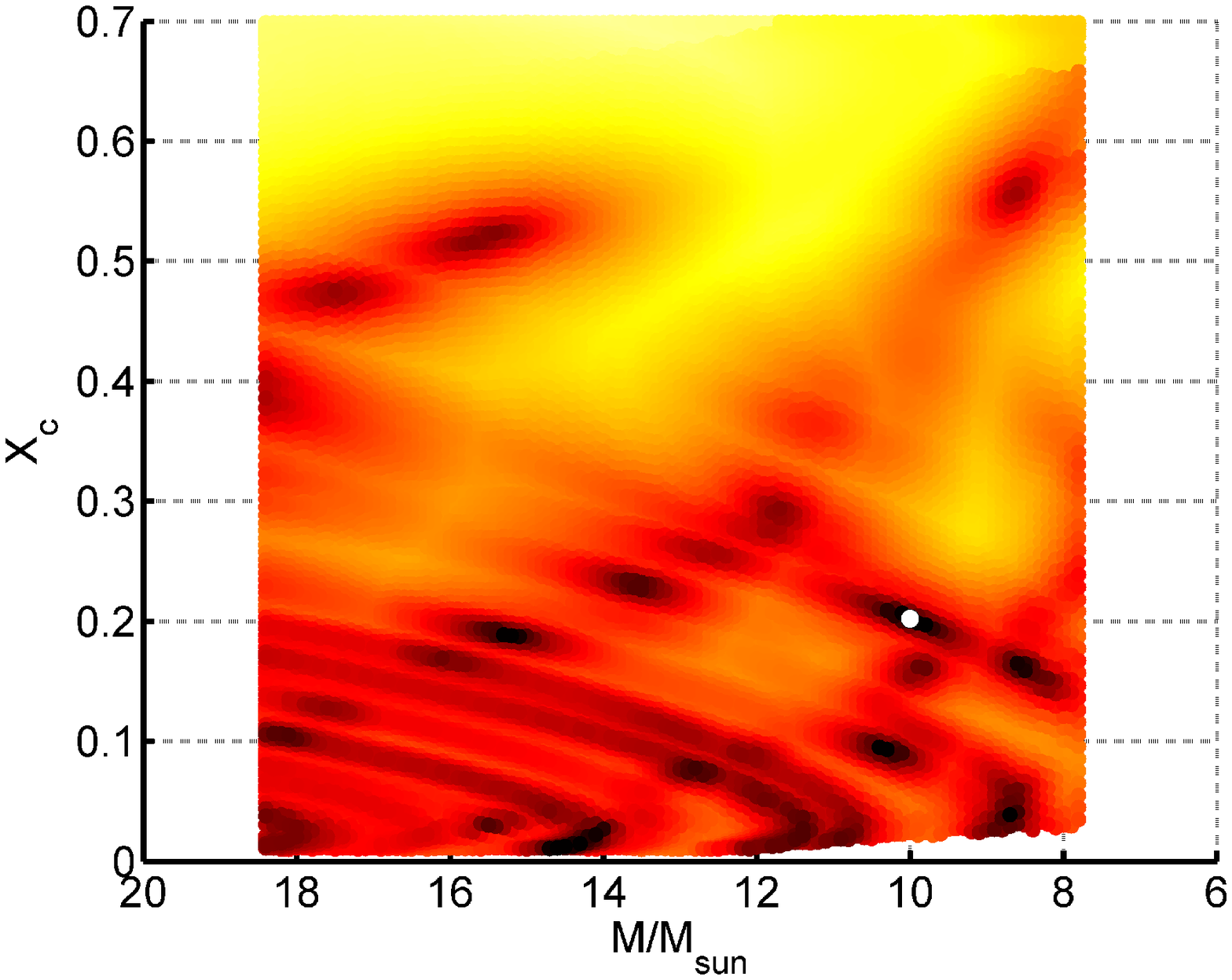}\includegraphics[width=0.5\textwidth]{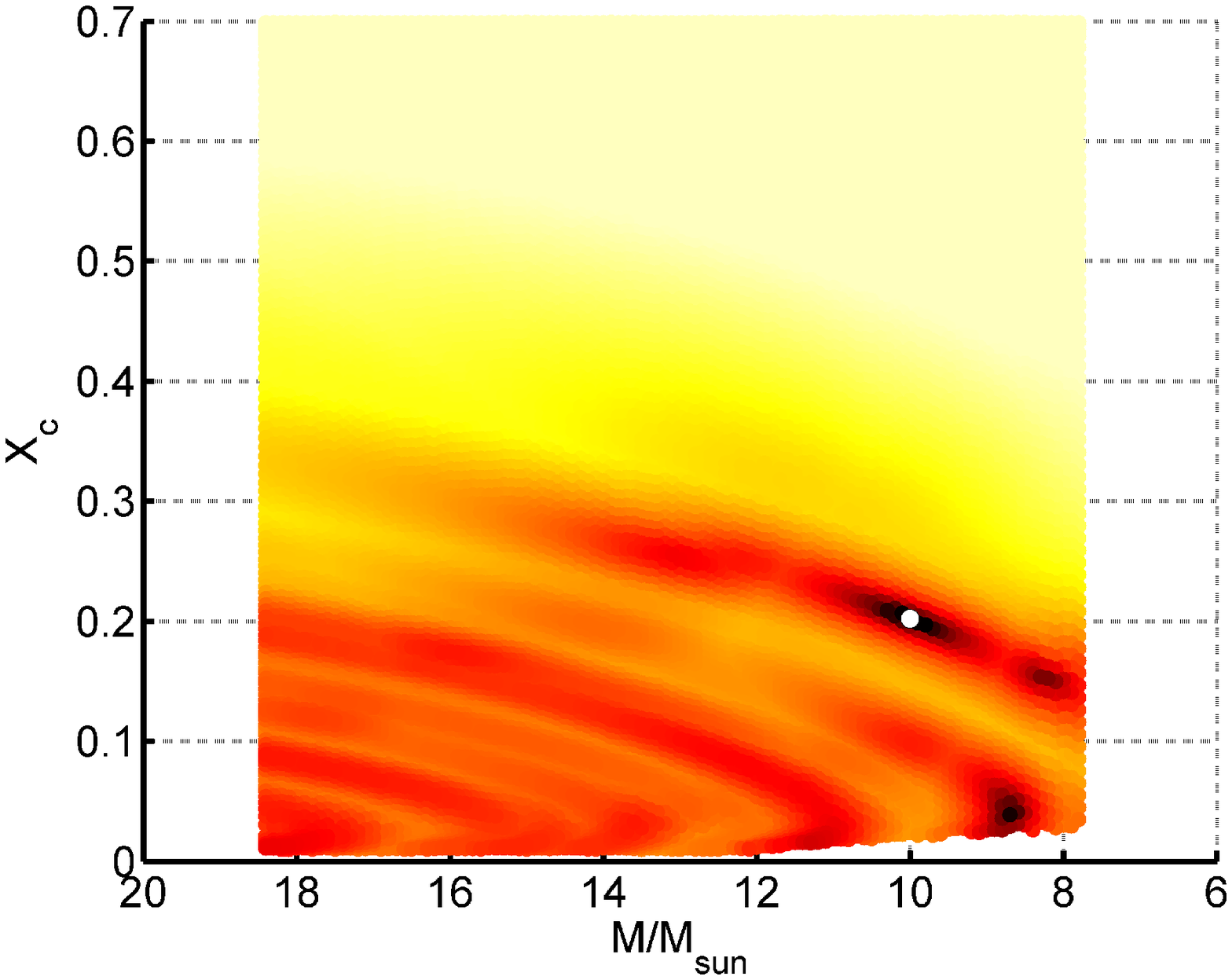}\vspace{0.1cm}
\includegraphics[width=0.7\textwidth]{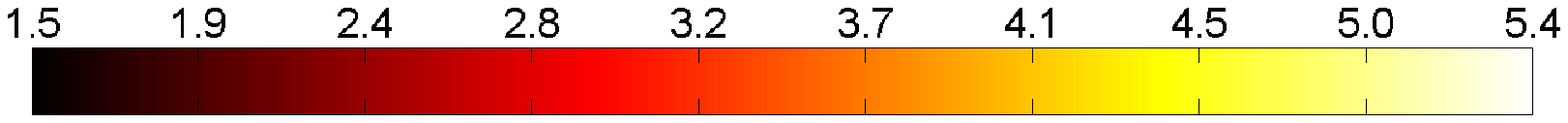}}
\caption{Colour-coded $\chi^2$ (logarithmic scale) in three different planes: $\log{T_{\rm eff}}$-$\log{L/L_\odot}$ (\textit{upper panels}), $\log{T_{\rm eff}}$-mean density (\textit{middle panels}), and mass-$X_{\rm c}$ (\textit{bottom panels}). White dots represent the position of the input model M0 in each parameter space. Right panels show $\chi^2$ computed including the identification of $\ell$ as a constraint, whereas in left panels $\ell$ is considered unknown ($\ell$= 0,1 or 2).}
\label{fig:chi2}
\end{center}
\end{figure}

We then compute the merit function on a sub-grid of models of {\sc betadat}, where $\alpha_{\rm OV}$ and $Z$ are the same as in M0. The behaviour of $\chi^2$ for main-sequence models of different mass and evolutionary status is shown in Fig. \ref{fig:chi2}.
In the case where $\ell$ is given as an observational constraint for all the modes (right panels), the properties and parameters of the input model M0 are easily recovered due to the appearance of an isolated global minimum in the $\chi^2$ function.
The increase of the number of $\chi^2$ minima in left panels ($\ell$ is unknown) compared to right panels allows to estimate the loss of information when mode identification is not available.
Nevertheless, even if no mode identification is available, the frequencies of 4 modes allow to constrain the parameter space in regions close to $\chi^2$ minima.

The results of this simple test show that the approach presented here is viable tool to determine the number and precision of observational data required to constrain the properties of $\beta$ Cephei pulsators. However, an extensive and thorough study is needed to investigate the effect of considering both additional uncertainties (e.g. on parameters such as overshooting and initial chemical composition, on rotational splittings and identification of $m$) as well as other constraints (e.g. luminosity, $T_{\rm eff}$, $\log{g}$, photospheric abundances, non-adiabatic constraints).

\acknowledgments{A.M. and J.M. acknowledge financial support from the Prodex-ESA Contract Prodex 8 COROT (C90199). A.M. is a \emph{Charg\'e de Recherches} and A.T. is a \emph{Chercheur qualifi\'e} of the FRS-FNRS.
}
\References{
\rfr Brassard, P., Fontaine, G., Bill\`eres, M., et al. 2001, ApJ, 563, 1013
\rfr Charpinet, S., Fontaine, G., Brassard, P., et al. 2005, A\&A, 437, 575
\rfr Scuflaire, R., Th\'eado, S., Montalb\'an, J., et al. 2008a, ApSS, 316, 83
\rfr Scuflaire, R., Montalb\'an, J., Th\'eado, S., et al. 2008b, ApSS, 316, 149
\rfr Thirion, A. \& Thoul, A. 2006, ESA SP-1306, 383
}

\end{document}